\documentclass[twocolumn,prd,showpacs,10pt]{revtex4}
\usepackage{graphicx}% Include figure files
\usepackage{dcolumn}% Align table columns on decimal point
\usepackage{bm}% bold math

\newcommand{\square}{\kern1pt\vbox{\hrule height 1.2pt\hbox{\vrule
width 1.2pt\hskip 3pt
\vbox{\vskip 6pt}\hskip 3pt\vrule width 0.6pt}\hrule
height 0.6pt}\kern1pt}
\newcommand{\beq}{\begin{equation}}
\newcommand{\beqn}{\begin{eqnarray}}
\newcommand{\eeq}{\end{equation}}
\newcommand{\eeqn}{\end{eqnarray}}

\begin{document}

\title{Correspondence between Loop-inspired and Braneworld Cosmology}
\author{Edmund J.~Copeland$^1$, James E. Lidsey$^2$
and Shuntaro Mizuno$^3$}
\affiliation{$^1$ School of Physics and Astronomy, University of Nottingham,
University Park, Nottingham, NG7 2RD, United Kingdom}
\affiliation{$^2$ Astronomy Unit, School of Mathematical Sciences, 
Queen Mary, University of London, Mile End Road,
London, E1 4NS, UK}
\affiliation{$^3$ Department of Physics, Waseda University, Okubo 3-4-1,
Shinjuku, Tokyo 169-8555, Japan}

\date{\today}

\begin{abstract}

Braneworld scenarios are motivated by string/M-theory and 
can be characterized by the way in which they modify the 
conventional Friedmann equations of Einstein gravity. An alternative
approach to quantum gravity, however, is
the loop quantum cosmology program. In the semi-classical limit,
the cosmic dynamics in this scenario can also be described by a set
of modified Friedmann equations. We
demonstrate that a dynamical correspondence can be established 
between these two paradigms at the level of the effective field 
equations. This allows qualitatively similar features
between the two approaches to be compared and contrasted as well as
providing a framework for viewing
braneworld scenarios in terms of constrained Hamiltonian systems.
As concrete examples of this correspondence, we illustrate the
relationships between different cosmological backgrounds representing
scaling solutions.
 
\end{abstract}

\vskip 1pc 
\maketitle
%\vskip 1pc

%

%========================================%
%<<<<<<<<<<<< SECTION I  >>>>>>>>>>>>>>%
%========================================%
\section{Introduction}

The high energy and high curvature regime of the very early universe 
provides a natural environment for  
investigating the cosmological consequences of quantum theories of
gravity.  At present, the two leading contenders for 
a theory of quantum gravity are string/M--theory and loop quantum 
gravity (LQG). (For recent reviews, see, e.g., Refs. 
\cite{stringreview,loopreview,bojreview}). 

Developments in string/M--theory over recent years have led to a 
new paradigm for early universe cosmology -- the braneworld scenario
(see \cite{branereview} for a review). 
In this picture, our observable, four--dimensional universe is 
viewed as a co--dimension one brane that is embedded in a five--
or higher--dimensional `bulk' space 
\cite{earlybranerefs,Arkani-Hamed:1998rs,R_S}. 
Propagation of the brane through 
the bulk is interpreted as cosmic expansion or contraction 
by an observer confined to the brane. The
brane--bulk dynamics typically results in modifications 
to the effective four--dimensional Friedmann equations of standard
cosmology. These can be parametrized in terms of deviations from the 
standard $H^2 \propto \rho$ behaviour and may become significant 
on either large or small scales, depending on the nature of 
the model under consideration. 

LQG, on the other hand, is a non--perturbative canonical quantization 
of Einstein gravity. Loop quantum cosmology (LQC) restricts the analysis of LQG to 
spatially homogeneous models \cite{boj1}. In the 
scenario developed by Bojowald \cite{boj2,boj3}, there are two 
critical scales, $a_i$ and $a_*$. Spacetime 
has a discrete structure below the scale $a_i \approx \ell_{\rm Pl}$, 
whereas classical gravity is recovered above $a_*$. The scale $a_*$ 
is sensitive to the quantization scheme adopted and can be significantly
larger than $a_i$. In this case, there exists a `semi--classical' 
phase in the history of the universe for $a_i < a \ll a_*$,
where spacetime may be viewed as a 
continuum but where non--perturbative quantum effects 
lead to modifications of the classical Friedmann equations. 
In particular, the kinetic energy of 
a scalar field becomes modified in such a way that $T=\dot{\phi}^2/D (a)$, 
where $D$ is some function of the scale factor that tends to 
unity for $a \gg a_*$. The cosmic dynamics of this phase 
has attracted considerable interest recently 
\cite{boj3,lqcpapers,mulryne}. 

To date, the braneworld and LQC scenarios have been investigated 
along separate lines. However, given that both paradigms are
motivated by quantum gravitational considerations, 
it is of interest to investigate the extent to which 
they may share any common features. This is the purpose of 
the present work. As a preliminary step in this 
direction, we consider braneworld and loop--inspired cosmologies where the 
matter sector is comprised of a self--interacting scalar field. 
We find that for a given Hubble expansion rate, the corrections 
to the field equations in both approaches can be directly related, thereby 
enabling the potentials of the scalar fields to also be compared. 
Furthermore, since LQC is based on a canonical quantization of gravity, 
such a link leads to an alternative description of 
braneworld cosmology in terms of a constrained Hamiltonian 
system. We illustrate this approach by focusing on 
cosmological solutions that exhibit scaling properties. Scaling solutions 
in cosmology are of particular importance because they provide insight 
into the asymptotic nature of a specific model together with 
its stability properties. They also provide a means 
of determining the generic behaviour of scalar fields in modified 
cosmology \cite{Copeland:2004qe,scalemodify}. 

The paper is organized as follows. In Section II we discuss 
how modifications to the Friedmann equations arise in a number of 
different settings and proceed 
to establish a general dynamical correspondence 
between braneworld and loop--inspired cosmologies in Section III. 
Section IV illustrates how scaling 
solutions in the Randall--Sundrum braneworld scenario
\cite{R_S} can be modeled in terms of a loop--inspired 
cosmology. In Section V we discuss the high--energy limits of a 
number of braneworld models where corrections 
to the Friedmann equations take a power law form. 
We conclude with a discussion in Section VI.

\section{Modified Cosmology}

\subsection{Conventional Cosmology}

The Einstein field equations for spatially 
isotropic Friedmann--Robertson--Walker (FRW) cosmologies 
sourced by a perfect fluid matter take the form 
\begin{eqnarray}
\label{standfriedmann}
H^2 = \frac{8 \pi \ell_{\rm Pl}^2}{3} \rho -\frac{k}{a^2} \\
\label{standconserve}
\dot{\rho} + 3H(\rho +p ) =0
\end{eqnarray}
where $H \equiv \dot{a}/a$ defines the Hubble parameter, $a(t)$ is the 
scale factor of the universe,
$\ell_{\rm Pl} = \sqrt{G\hbar}$ is the Planck length,
$\rho$ and $p$ denote the energy density 
and pressure of the fluid, a dot denotes 
differentiation with respect to cosmic time, and 
the constant $k$ takes values $ \{ -1, 0, +1 \}$
for negatively--curved, spatially flat, and positively--curved
universes, respectively. 
The dynamical properties of the matter source are determined through its 
equation of state: 
\begin{equation}
\label{standeos}
p=[ \gamma (\rho ) - 1 ] \rho
\end{equation}
where, in general,  the barotropic index $\gamma$ 
is an analytic function of the energy density, or equivalently, 
the scale factor. In principle, the evolution 
of the energy density is determined by integrating the 
conservation equation (\ref{standconserve}): 
\begin{equation}
\label{quadrho}
\rho (a) = \rho_i e^{-3\int_{a_i}^a d \ln a \, \gamma (a)}
\end{equation} 

The fluid may be modeled in terms of a scalar field, 
$\varphi$, minimally coupled to Einstein gravity and 
self--interacting through a potential, $V(\varphi )$, with 
energy density and pressure given by $\rho_{\varphi} = \dot{\varphi}^2/2
+V$ and $p_{\varphi}= \dot{\varphi}^2/2 -V$. The equation of state 
parameter then takes the 
form  $\gamma_{\varphi} = 2 \dot{\varphi}^2/(\dot{\varphi}^2
+2V)$ and the conservation equation (\ref{standconserve}) 
becomes 
\begin{equation}
\label{standscalareom}
\ddot{\varphi}+3H\dot{\varphi} + \frac{dV}{d\varphi} =0
\end{equation}

The `scaling' solution for such a field arises when 
the field's potential and kinetic energies scale at the 
same rate, $\dot{\varphi}^2 /V = {\rm constant}$. 
For a spatially flat FRW universe, this leads to 
the power--law solution, $a \propto t^{2/\lambda^2}$ driven by 
an exponential potential, $V \propto \exp (-  
\sqrt{8\pi \ell_{\rm Pl}^2} \lambda \varphi )$ \cite{lm}. 

\subsection{Braneworld Cosmology}

In this subsection we consider the 
class of spatially isotropic FRW world--volume metrics, 
where a canonical scalar field, $\chi$, that self--interacts through 
a potential, $W(\chi )$, is confined to the brane with 
energy density $\rho_{\chi} = 
\dot{\chi^2}/2 +W(\chi )$, pressure $p_{\chi}=\dot{\chi}^2/2-W(\chi )$
and equation of state, $\gamma_{\rm B} = 
2\dot{\chi}^2/(\dot{\chi}^2+2W)$.
The dynamics for a wide class of such braneworlds 
can be described in terms of a generalized Friedmann equation of 
the form \cite{Copeland:2004qe}
\begin{equation}
\label{branefriedmann}
H^2 = \frac{8 \pi \ell_{\rm Pl}^2}{3} \rho_{\chi} L^2(\rho_{\chi})
-\frac{k}{a^2}
\end{equation}
where modifications 
to conventional cosmology are determined by the form of 
the function $L(\rho_{\chi} )$. 
If there is no transfer of energy--momentum between the 
brane and bulk dimensions, the  standard conservation
equation (\ref{standconserve}) holds: 
\begin{equation}
\label{braneconserve}
\dot{\rho}_{\chi} + 3H(\rho_{\chi} +p_{\chi} )=0
\end{equation}

Eqs. (\ref{branefriedmann}) and (\ref{braneconserve}) 
are sufficient to completely determine the dynamics once the potential 
of the scalar field has been specified. Each 
braneworld model is therefore characterized by 
the dependence of the function $L(\rho_{\chi} )$ on the energy density. 
For example, $L = \sqrt{1+(\rho_{\chi}/2\sigma)}$ 
in the Randall--Sundrum type II 
scenario \cite{R_S}, where $\sigma$ represents the tension of the brane
\cite{cline,SMS,Binetruy};
$L = \sqrt{1-(\rho_{\chi}/2|\sigma|)}$ in the
Shtanov-Sahni model \cite{Shtanov:2002mb}; and 
$L = (1/\sqrt{D\rho_{\chi}})[\mp 1 + \sqrt{1+D\rho_{\chi}}]$ in 
the Dvali-Gabadadze-Porrati scenario \cite{Dvali:2000hr}. 
Another class of modified cosmologies are the Cardassian models, 
where $L = \sqrt{1+B \rho_{\chi}^n}$ for some positive constant $B$ 
and $n<-1/3$ \cite{Freese:2002sq}. A power--law form 
for the correction function $L^2 \propto \rho_{\chi}^q$ corresponds to 
the high--energy limit of a number of braneworld scenarios, 
including the Randall--Sundrum scenario $(q=1)$ and the extension of 
the Randall--Sundrum scenario to include a Gauss--Bonnet 
combination of curvature invariants in the 
five--dimensional bulk action. In this case, $L^2\propto 
\rho_{\chi}^{-1/3}$ \cite{gbbrane}. Effective Friedmann equations 
with a power--law correction to the energy density \cite{chung}
are also possible 
in models based on Ho\v{r}ava--Witten theory \cite{HW} compactified 
on a Calabi--Yau three--fold. 
 
Eqs. (\ref{branefriedmann}) and (\ref{braneconserve}) may  
be written in the standard form of the FRW Einstein field equations 
sourced by a perfect fluid with an effective energy density and pressure:
\begin{eqnarray}
\label{rhoeffB}
\rho_{\rm B, eff} \equiv \rho^{n (\rho_{\chi} )}_{\chi} 
\\
p_{\rm B, eff} = (\gamma_{\rm B, eff} - 1) \rho_{\rm B, eff}
\end{eqnarray}
where the equation of state parameter is defined by 
\begin{equation}
\label{gammaeffB}
\gamma_{\rm B,eff} \equiv -\frac{1}{3} \frac{d (n \ln \rho_{\chi} )}{d\ln a}  
\end{equation}
and we have introduced an index $n$ defined as 
\begin{equation}
\label{defn}
n(\rho_{\chi}) \equiv 1+2 \frac{\ln L(\rho_{\chi} )}{\ln \rho_{\chi}}
\end{equation}
In general, this is a function of the energy density on the brane and 
its form defines the braneworld model.  
Eq. (\ref{gammaeffB}) follows by differentiating 
Eq. (\ref{rhoeffB}) and substituting in the standard 
conservation equations (\ref{standconserve}) and 
(\ref{braneconserve}). This implies that
\begin{eqnarray}
H^2_{\rm B,eff} = \frac{8 \pi \ell_{\rm Pl}^2}{3} \rho_{\rm B,eff}
- \frac{k}{a^2} 
\\
\dot{\rho}_{\rm B,eff} =-3H(\rho_{\rm B,eff}
+ p_{\rm B,eff} )
\end{eqnarray}

\subsection{Loop-Inspired Cosmology}

The standard FRW equations (\ref{standfriedmann}) and (\ref{standscalareom}) 
may be derived by viewing the dynamics as a constrained 
Hamiltonian system ${\cal{H}} =0$, where the 
Hamiltonian is given by 
\begin{equation}
\label{classham}
{\cal{H}} = - \frac{3}{8\pi \ell_{\rm Pl}^2} \left( \dot{a}^2 
+k^2 \right) a + \frac{1}{2} a^{-3} p_{\varphi}^2 +a^3 V(\varphi ) 
\end{equation}
and $p_{\varphi} = a^3\dot{\varphi}$ is the momentum 
canonically conjugate to the scalar field. 

An alternative way of considering modified FRW cosmologies is to introduce
corrections to the Hamiltonian (\ref{classham}). For example, 
we may consider the class of modified cosmologies where the inverse 
volume factor of the momentum of a scalar field, $\phi$, is 
replaced by a function $D(a)$ of the scale factor such that: 
\begin{equation}
\label{semiclassham}
{\cal{H}}_{\rm mod} = -\frac{3}{8\pi \ell^2_{\rm Pl}} 
\left( \dot{a}^2 +k^2 \right) a
+ \frac{1}{2} D (a) a^{-3} p^2_{\phi} +a^3V =0  
\end{equation}
where the conjugate momentum 
is now given by $p_{\phi} = D^{-1}(a) a^3 \dot{\phi}$. 

The primary motivation for considering models of this type 
comes from LQC. In isotropic LQC, the classical variables 
are the triad component $|p|=a^2$ and the connection 
component $c=\frac{1}{2}(k-\gamma \dot{a})$, where 
$\gamma \approx 0.274$ is the Barbero--Immirzi parameter \cite{bipara}. 
The divergent geometrical density,  $a^{-3}$, that is 
present in the classical Hamiltonian (\ref{classham}) 
is quantized by employing the classical 
identity $a^{-3} = [3(8\pi \ell_{\rm Pl}^2 l)^{-1} \{ c, 
|p|^l \} ^{3/(2-2l)}$, where $\{ A,B \}$ denotes the Poisson bracket and
$l$ is a constant, and replacing the connection with holonomies
along curves in space associated with the connection and 
triad \cite{bojreview}. In this case, it can be shown that 
inverse factors of the scale factor are not required if $0 <l <1$ 
\cite{bojreview}. 
Consequently, quantization results in a descrete spectrum of eigenvalues
for the geometrical density operator that remains {\em bounded} 
as the spatial volume of the universe vanishes. 
This spectrum is closely approximated by a 
continuous `eigenvalue function' of the scale factor  
$d_{j,l}(a) \equiv D(q)a^{-3}$, where \cite{bojreview,boj2}
\begin{eqnarray}
\label{defD}
 D(q) =
\left\{\frac{3}{2l}q^{1-l}\left[(l+2)^{-1}
\left((q+1)^{l+2}-|q-1|^{l+2}\right)\right.\right. \nonumber\\
 - \left.\left.\frac{1}{1 + l}q
\left((q+1)^{l+1}-{\rm
sgn}(q-1) 
|q-1|^{l+1}\right)\right]\right\}^{3/(2-2l)} 
\end{eqnarray}
and $q=a^2/a_*^2$, $a_*^2= a_i^2 j/3$, $a_i= \sqrt{\gamma} 
\ell_{\rm Pl}$. 
The parameter $j$ must take half--integer (positive) values, but 
is otherwise arbitrary \cite{amigpara}. 
It arises because there is an ambiguity in expressing the classical geometrical 
density in terms of holonomies, since any irreducible 
${\rm SU}(2)$ representation with spin $j$ may be chosen. 
In the limit $a_i < a \ll a_*$, 
the eigenvalue function asymptotes to a power--law: 
\begin{equation}
\label{approxD} 
 D (a) \propto
(3/(1+l))^{3/(2-2l)}(a/a_*)^{3(2-l)/(1-l)} 
\end{equation}
and classical behaviour, corresponding to $D \rightarrow 1$, 
is recovered for $a > a_*$. The eigenvalue 
function is peaked around $a\approx a_*$.
The semi--classical phase of LQC corresponds to the 
regime where $a_i<a<a_*$, and the cosmic dynamics 
during this phase can be determined from an effective 
Hamiltonian (\ref{semiclassham}), where the function 
$D (a)$ is given by Eq. (\ref{defD}) for $j> 3/2$ and $0<l<1$ 
\cite{boj2,boj3,bojvand}.
(The value of $j$ determines the duration of this phase).  

The above discussion therefore motivates us to consider cosmological 
models defined by the effective Hamiltonian 
(\ref{semiclassham}), where $D$ may be viewed as an unspecified function 
of the scale factor, in the same way that the potential 
of the scalar field is regarded as a free function that is ultimately determined 
by particle physics considerations. We will refer to $D(a)$ as the `kinetic
correction' function. 
The field equations for this scenario then follow directly from 
the Hamiltonian constraint ${\cal{H}}_{\rm mod} =0$ and Hamilton's equations 
and are given by  
\beqn
H^2 &=&\frac{8 \pi l_{Pl}^2}{3}
\left[\frac{1}{2D(a)} \dot{\phi}^2 + V \right] - \frac{k}{a^2},
\label{loopfriedmann}
\\
%\frac{\ddot{a}}{a} &=&-\frac{8 \pi l_{Pl}^2}{3}
%\left[\frac{\dot{\phi}^2}{D}
%\left(1-\frac{3\alpha}{4}\right) -V\right],\\
\label{loopscalareom}
\ddot{\phi} &+& 3H \left( 1- \frac{1}{3} \frac{d \ln D}{d \ln a} 
\right) \dot{\phi} + D \frac{d V}{d \phi}=0 .
\eeqn

Eqs. (\ref{loopfriedmann}) and (\ref{loopscalareom}) 
may also be expressed in the form of 
a conventional cosmology (\ref{standfriedmann})--(\ref{standeos}) by 
defining an effective energy and pressure \cite{mulryne}: 
\beqn
\label{rhoeffL}
\rho_{\rm L,eff} &=& \frac{\dot{\phi}^2}{2D} + V,
\\
\label{peffL}
p_{\rm L,eff} &=&  \frac{\dot{\phi}^2}{2D} 
\left( 1- \frac{1}{3} \frac{d \ln D}{d \ln a} \right) - V .
\eeqn
It follows that Eqs. (\ref{loopfriedmann}) and (\ref{loopscalareom}) 
transform to 
\begin{eqnarray}
H^2_{\rm L,eff} = \frac{8 \pi \ell_{\rm Pl}^2}{3} \rho_{\rm L,eff}
- \frac{k}{a^2} 
\\
\dot{\rho}_{\rm L,eff} =-3H(\rho_{\rm L,eff}
+ p_{\rm L,eff} )
\end{eqnarray}
where the effective equation of state takes the form 
\begin{equation}
\label{gammaeffL}
\gamma_{\rm L,eff} = \frac{2\dot{\phi}^2}{\dot{\phi}^2 +2DV}
\left( 1-\frac{1}{6} \frac{d\ln D}{d\ln a} \right)
\end{equation}

\section{Correspondence between Loop--Inspired and Braneworld Cosmology}

The correspondence between the braneworld and loop--inspired 
descriptions of cosmic dynamics is now established by identifying the 
effective equations of state for the two scenarios, i.e., 
$\gamma_{\rm B,eff}=\gamma_{\rm L,eff}$. Equating 
Eqs. (\ref{gammaeffB}) and (\ref{gammaeffL}) therefore leads to 
the constraint equation: 
\begin{equation}
\label{correspondence}
\frac{2\dot{\phi}^2}{\dot{\phi}^2 +2DV}
\left( 1-\frac{1}{6} \frac{d\ln D}{d\ln a} \right) = 
-\frac{1}{3} \frac{d}{d\ln a} (n \ln \rho )
\end{equation}
where for notational ease we drop the subscript 
$\chi$ on the brane energy density in what follows. 
Imposing the constraint (\ref{correspondence})
implies that the two different scenarios lead to cosmologies with  
identical Hubble parameters, $H_L =H_B =H$, when these are viewed as 
functions of cosmic time, or equivalently, as functions of the scale 
factor. The correspondence is not one--to--one since 
the loop--inspired description has two free functions --  the potential 
$V(\phi)$ and the function $D(a)$. Thus, a given 
braneworld model will correspond to a class of loop--inspired 
cosmologies and vice--versa. More specifically, let us 
suppose that a particular braneworld scenario
has been developed such that the FRW equations 
(\ref{branefriedmann}) and (\ref{braneconserve}) 
have been solved for a given form of $n(\rho )$ 
to determine the equation of state for 
the scalar field, $\gamma_{\rm B}  =\gamma_{\rm B}  (a)$, 
and therefore the evolution of the energy density $\rho  (a)$.
Thus, the right--hand side of Eq. (\ref{correspondence}) 
is in principle known. A natural way of characterizing the classes of 
loop--inspired scenarios that lead to such cosmic behaviour is 
to specify the ratio of the 
field's potential and kinetic energies as a function 
of the scale factor: $V/\dot{\phi}^2 \equiv f(a)$. 
Eq. (\ref{correspondence}) can then be expressed in the 
form of a non--linear Bernoulli equation:
\begin{equation}
\label{genbern}
\frac{dD}{d\ln a} - \left[ 6+\frac{d(n\ln \rho )}{d\ln a} \right] D 
= 2f(a)\frac{d(n\ln \rho )}{d \ln a} D^2
\end{equation}
and Eq. (\ref{genbern}) may be solved in terms of a single 
quadrature by defining a new variable $D\equiv G^{-1}$. We find that 
\begin{equation}
\label{genbernsol}
D^{-1} = \frac{1}{a^6\rho^n} \left[ C-2\int d(\rho^n ) \, f(a) a^6
\right]
\end{equation}
where $C$ is an integration constant. 

The question that now arises is to identify appropriate 
forms for the ratio $f(a)$. In conventional cosmology, 
scaling solutions where 
the field's potential and kinetic energies redshift at the same rate
have
played an important role. This latter constraint could 
be invoked to characterize scaling solutions 
in loop--inspired cosmology such that $f=V/\dot{\phi}^2 ={\rm
constant}$.
Alternatively, inspection of the effective 
energy density (\ref{rhoeffL}) suggests that the 
restriction $f(a) = D(a)$ 
also represents a scaling property since in this 
case it is the ratio of the field's {\em effective} kinetic energy, 
$\dot{\phi}^2/D$, to its potential energy that remains fixed. 
In view of this, we now consider each of these possibilities in turn. 

\subsection{Case A: $DV/\dot{\phi}^2 = \beta = {\rm constant}$}

In this 
case, Eq. (\ref{correspondence}) may be integrated directly to yield 
the form of the correction function $D$ in loop--inspired cosmology 
directly in terms of the modification to the Friedmann equation, 
$n(\rho )$, in the braneworld cosmology: 
\begin{equation}
\label{caseAcorr}
\frac{D}{D_c} = a^6 \rho^{(1+2\beta )n(\rho)}
\end{equation}
where $D_c$ is a constant determined by the integration constant.
Hence, if a particular braneworld cosmology is known, so that 
$\rho (a)$ is determined as a function of the scale factor, 
the form of $D (a) $ can be deduced immediately. In particular, 
a braneworld scaling cosmology, characterized by constant $\gamma_{\rm B}$
has $\rho \propto a^{-3\gamma_{\rm B}}$. 

The dependence of the potential on the scale factor follows
immediately from the Friedmann equation (\ref{loopfriedmann}): 
\begin{equation}
\label{potA}
V(a) = \frac{2\beta}{1+2\beta} \rho^{n(\rho )}
\end{equation}
Employing the identity $\dot{\phi}=Hd\phi/d\ln a$ then yields 
the dependence of the scalar field on the size of the universe 
up to a single quadrature: 
\begin{equation}
\label{fieldA}
\phi (a) =\pm \sqrt{\frac{3D_c}{4\pi \ell_{\rm Pl}^2
(1+2\beta )}} \int da \, a^2 
[\rho (a)]^{(1+2\beta)n(a)/2}
\end{equation}

Evaluating the integral (\ref{fieldA}) 
and inverting the result then yields, in
principle, the scalar field potential $V(\phi )$ from Eq. (\ref{potA}). 
It is surprising that given a braneworld cosmology, one need only  
perform a single integral, Eq. (\ref{fieldA}), in order to 
determine the corresponding form of the scalar field potential in
the loop--inspired cosmology. Furthermore, note that 
the scalar field expression (\ref{fieldA}) has no 
explicit dependence on the Hubble parameter or the 
kinetic correction function $D$. On the other hand, 
for the field to have positive kinetic energy 
requires that $D_c(1+2 \beta ) >0$. If this condition 
is not satisfied, the field behaves effectively as 
a phantom matter source where the null energy condition 
is violated.

\subsection{Case B: $V/\dot{\phi}^2 = \epsilon = {\rm constant}$}
 
In this case, Eq. (\ref{correspondence}) reduces to 
a non--linear Bernoulli equation in the dependent variable $D(a)$: 
\begin{equation}
\label{bernoulli}
\frac{dD}{d\ln a} - \left[ 6+\frac{d}{d\ln a}(n\ln \rho ) \right]
D = \left[ 2\epsilon \frac{d}{d \ln a} (n \ln \rho ) \right] 
D^2_l
\end{equation}
which admits an integral in terms of 
a single quadrature: 
\begin{equation}
\label{bernsol}
D^{-1}_l (a) = -2\epsilon + \frac{1}{a^6\rho^{n(\rho )}} 
\left[ C+12\epsilon \int da \, a^5 \rho^{n(\rho )} \right] 
\end{equation}

The Friedmann equation (\ref{loopfriedmann}) leads to the potential 
\begin{equation}
\label{potB}
V(a) = 2\epsilon a^6 \rho^{2n} \left[ C+12\epsilon \int da \, a^5 \rho^n
\right]^{-1}
\end{equation}
whereas the scalar field is given by 
\begin{equation}
\label{fieldB}
\phi (a) = \pm \sqrt{\frac{3}{4\pi \ell_{\rm Pl}^2}} \int 
d \ln a \, \sqrt{\frac{D(a)}{1+2\epsilon D(a)}}
\end{equation}
or, equivalently, after substitution of the solution (\ref{bernsol}), by
\begin{equation}
\label{equivfieldB}
\phi (a) =\pm \sqrt{\frac{3}{4\pi \ell_{\rm Pl}^2}} \int da \,
a^2\rho^{n/2}
\left[ C+12\epsilon \int da \, a^5 \rho^n \right]^{-1/2}
\end{equation}

For $D>0$, the positivity of the scalar field kinetic 
energy implies that $\epsilon$ is constrained such that $\epsilon > -1/(2 D)$.
Moreover, we see once more that the integrand (\ref{equivfieldB}) can be 
expressed directly in terms of the known evolution of the energy density
on the brane.

To summarize thus far, the above correspondences provide an equivalent 
means of discussing braneworld scenarios in terms of 
the modified constrained Hamiltonian system (\ref{semiclassham})
for arbitrary equations of state 
$\{ \gamma_{\rm B},  \gamma_{\rm L} \}$. 
In the following two Sections, we develop the correspondence further 
for specific braneworld scenarios. 

\section{Randall--Sundrum Braneworld as a Loop--Inspired Cosmology}

In \cite{Copeland:2004qe}, we established a unified framework 
for determining the general form of the scalar field potential $W(\chi )$ 
that leads to an attractor scaling solution 
in a given braneworld scenario. The form of the 
potential is related to the form of $n(\rho )$ defined in Eq. (\ref{defn}). 
In this Section, we consider the scaling solutions that arise in 
the Randall--Sundrum type II scenario and illustrate 
how these models can be effectively described in terms of loop--inspired 
cosmologies. 

In general, a braneworld scaling solution has a constant equation
of state on the brane, $\gamma_{\rm B}={\rm constant}$, although 
it is important to note that this does not necessarily imply that 
the effective equation of state, $\gamma_{\rm B,eff}$, is also constant.  
The conservation equation (\ref{braneconserve}) may 
be integrated immediately to yield $\rho = a^{-3\gamma_{\rm B}}$, 
where the integration constant has 
been absorbed by an appropriate rescaling, without loss of generality. 
The effective energy density (\ref{rhoeffB}) for the 
Randall--Sundrum model is therefore given by 
\begin{equation}
\label{RSeff}
\rho^n = \frac{1}{a^{3\gamma_{\rm B}}} + 
\frac{1}{2\sigma a^{6\gamma_{\rm B}}}
\end{equation}
where $\sigma$ represents the brane tension. 

In the following, we consider only spatially 
flat cosmologies ($k=0$). In this case,  
the scalar field potential that drives this scaling solution 
has a hyperbolic form \cite{Hawkins:2000dq}: 
\begin{equation}
\label{hawklid}
W(\chi ) = \sigma (2-\gamma_{\rm B} ) {\rm cosech}^2 \, 
\left( \sqrt{6\pi \ell_{\rm Pl}^2 \gamma_{\rm B}} \chi \right)
\end{equation}

\subsection{Case A: $DV/\dot{\phi}^2 = \beta = {\rm constant}$}

For this case, the kinetic correction function is given by 
Eq. (\ref{caseAcorr}) after substitution of Eq. (\ref{RSeff}): 
\begin{equation}
\label{Dform}
D(a)=\frac{D_c}{(2\sigma)^{1+2\beta}} 
a^{6[1-\gamma_{\rm B}( 1 +2\beta)]} 
\left[ 1+2\sigma a^{3 \gamma_{\rm B}} \right]^{1+2\beta}
\end{equation}
and the evolution of the scalar field is then determined 
from Eq. (\ref{fieldA}): 
\begin{equation}
\label{genfield}
\phi (a) = \phi_0 \int dx \, x^p (1+x)^q
\end{equation}
where $p\equiv [1-2\gamma_{\rm B}(1+\beta )]/\gamma_{\rm B}$,
$q \equiv (1+2\beta )/2$, $x \equiv 2\sigma a^{3\gamma_{\rm B}}$ and we 
have absorbed all constants into the constant $\phi_0$ for notational 
simplicity. For any integer $p$, 
Eq. (\ref{genfield}) may be evaluated by integrating 
by parts a sufficient number of times, although the result is 
not necessarily invertible. 
More generally, Eq. (\ref{genfield}) may be 
evaluated to yield the solution 
\begin{equation}
\label{hyperfield}
\phi (a) = \frac{\phi_0}{(1+p)} x^{1+p} {_2}F_1 (1+p, -q, 
2+p, -x)
\end{equation}
where ${_2}F_1$ is the hypergeometric function. 

%We are free to choose any value of the parameter $\beta$ 
%satisfying $\beta >-1/2$. Indeed,  

An invertible example is found for $\beta = 
(1-2\gamma_{\rm B})/(2\gamma_{\rm B})$ ($p=0$). 
It follows from Eq. (\ref{genfield}) that 
\begin{equation}
\label{choosebeta}
\phi (a) = \frac{2\gamma_{\rm B} \phi_0}{1+\gamma_{\rm B}} \left[
\left( 1+ 2\sigma a^{3\gamma_{\rm B}} \right)^{(1+\gamma_{\rm B})/
(2\gamma_{\rm B})} -1 \right]
\end{equation}
and the potential in this case is determined immediately 
from Eq. (\ref{potA}). We find that it is given in terms 
of powers of the scalar field: 
\begin{eqnarray}
\label{potchoosebeta}
V(\phi )  = 2\sigma \left( \frac{1-2\gamma_{\rm B}}{1-\gamma_{\rm B}}
\right) \left(1+ \frac{\phi}{\tilde{\phi}_0} 
\right)^{(2\gamma_{\rm B})/(1+\gamma_{\rm B})}
\nonumber \\
\times
\left[ \left(1+ \frac{\phi}{\tilde{\phi}_0} 
\right)^{(2\gamma_{\rm B})/(1+\gamma_{\rm B})} -1 \right]^{-2}
\end{eqnarray}
where $\tilde{\phi}_0= 2\gamma_{\rm B}\phi_0/(1+\gamma_{\rm B})$.
This yields a vanishing potential for $\gamma_{\rm B} =1/2$ and 
the field has positive kinetic energy 
for $0<\gamma_{\rm B} <1$ ($\beta >-1/2$).

It follows from the properties of the hypergeometric function that a 
second invertible solution is possible for $( p,q)=(-1/2,-1)$. 
This corresponds to $\beta =-3/2$ and a particular choice 
for the equation of state on the brane, $\gamma_{\rm B} = -2/3$. 
This latter value corresponds to the equation of state 
for a universe dominated by a gas of domain walls and therefore has 
interesting physical consequences. Since $\beta < -1/2$, the field 
will represent an effective phantom field if $D_c>0$.  
The scalar field and potential energy evolve as 
\begin{eqnarray}
\phi (a) = - 2\phi_0 {\rm tan}^{-1} \left( \sqrt{2\sigma} a^{3\gamma_{\rm
B}/2}
\right) 
\\
V(\phi ) = 3\sigma \cos^2 \left( \frac{\phi}{2\phi_0} \right)
{\rm cosec}^4 \left( \frac{\phi}{2\phi_0} \right)
\end{eqnarray}

A third invertible solution follows from the 
identity ${_2}F_1 (1,1; 2, z) = - z^{-1} \ln(1-z)$. 
For this case, we have $(p,q)=(0,-1)$, which implies that 
$\beta = -3/2$ and $\gamma_{\rm B} = -1$, corresponding to a fluid 
on the brane well into the phantom regime. The solutions in this case are 
\begin{equation}
\phi(a) = \phi_0 \ln \left (1+  2\sigma a^{3\gamma_{\rm B}} \right)
\end{equation}
and 
\begin{equation}
%V(\phi) = {3 \sigma e^{\frac{\phi}{\phi_0}} \over (e^{\frac{\phi}{\phi_0}} -1)^2}.
\label{thirdpot}
V = \frac{3\sigma}{4} {\rm cosech}^2 \, \left( 
\frac{\phi}{2\phi_0} \right)
\end{equation}
It is interesting that the form of the potential (\ref{thirdpot}) 
for the scalar field in the loop-inspired scenario is qualitatively 
similar to the potential (\ref{hawklid}) that drives a scaling solution 
for a non-phantom field on the Randall-Sundrum brane. 
Hence, Eq. (\ref{thirdpot}) has the interesting limiting behaviour of 
$V \propto 1/\phi^2$ as $\phi \to 0$, corresponding to 
the high-energy limit $(a \rightarrow 0)$.

\subsection{Case B: $V/\dot{\phi}^2 = \epsilon = {\rm constant}$}
 
This case is more algebraically involved. 
Nonetheless, we are able to determine the form of the correction
function 
directly by evaluating Eq. (\ref{bernsol}). It follows that 
for $\gamma_{\rm B} \ne 1$, 
\begin{eqnarray}
\label{DsolcaseB}
\frac{1}{D} &=& 2\sigma C \frac{a^{6(\gamma_{\rm B}-1)}}{1+2\sigma
a^{3\gamma_{\rm B}}} 
\nonumber \\
&+& \frac{2\gamma_{\rm B}\epsilon}{1-\gamma_{\rm B}}
\frac{1}{1+2\sigma a^{3\gamma_{\rm B}}} \left[ 
1+ \left( \frac{1-\gamma_{\rm B}}{1-\frac{1}{2}\gamma_{\rm B}}
\right) \sigma a^{3 \gamma_{\rm B}} \right]\nonumber\\
\end{eqnarray}
For this form of the kinetic correction function, the 
integration constant $C$ plays an important role. If $C=0$, the 
correction function becomes singular at a finite value of the scale factor 
if $\gamma_{\rm B} >1$. Moreover, if $C\ne 0$, a decaying 
power--law behaviour is recovered at high energies (small 
values of the scale factor) for $0<\gamma_{\rm B} <1$, 
in agreement with the asymptotic behaviour of the eigenvalue 
correction function (\ref{approxD}) of LQC.  

For completeness, the solution for $\gamma_B =1$ can also 
be determined and is given by
\begin{eqnarray}
\label{DsolcaseB2}
\frac{1}{D} = \frac{4\epsilon (3 \ln a + \sigma a^3 - \frac{1}{2}) + 2\sigma C) }{1+2 \sigma a^3}.
\end{eqnarray}

\section{Power-law corrections inducing braneworlds 
as loop-inspired cosmologies}

A number of braneworld scenarios receive a power--law correction 
to the Friedmann equation of the 
form $L^2 \propto \rho^q$ in the high energy limit. 
This asymptotic behaviour includes the Randall--Sundrum scenario
($q=1$) \cite{cline,SMS,Binetruy}
and the extension of this model to include
a Gauss-Bonnet term in the five--dimensional action ($q=-1/3$) 
\cite{gbbrane}.
In this Section, we consider this class of models with 
arbitrary $q$ and investigate the spatially flat 
scaling solutions where $0 < \gamma_{\rm B} < 2$ is constant. 
(We will also assume implicitly that the kinetic correction 
function $D>0$ and $\beta >-1/2$ so that the 
effective field $\phi$ is a non-phantom field, although we note that this assumption can be relaxed). 
The effective energy density (\ref{rhoeffB}) is therefore given by 
\begin{equation}
\rho^n = A a^{-3\gamma_{\rm B}(q+1)},
\label{rho_to_n_pcbc_g}
\end{equation}
where $A$ is an arbitrary constant with dimension $m^{-4q}$.
The scalar field potential that drives this scaling solution has
an inverse power-law form \cite{Hawkins:2000dq,Mizuno:2004xj}:
\begin{equation}
\label{powerchi}
W(\chi ) = \left(1-\frac{\gamma_\chi}{2}\right)
\left(6\pi A q^2 \ell_{\rm Pl}^2 \gamma_\chi \right)^{-\frac{1}{q}}
\chi^{-\frac{2}{q}}
\end{equation}

It follows from Eq.~(\ref{gammaeffB}) that the 
effective equation of state on the brane is 
a constant, with numerical value 
\begin{equation}
\gamma_{\rm B,eff} = \gamma_{\rm B} (q+1)
\end{equation}
implying that the effective equation of state
in the corresponding loop-inspired cosmology is also constant,  
$\gamma_{\rm L,eff} = \gamma_{\rm B} (q+1)$. This 
property implies that the dependence of the potential on the scale factor 
in the loop--inspired model can be determined directly 
from the Friedmann equation (\ref{loopfriedmann}):
\begin{equation}
V(a) = \frac{2-\frac{1}{3} \frac{d \ln D}{d \ln a} - \gamma_{\rm B} (q+1)}
{2-\frac{1}{3} \frac{d \ln D}{d \ln a}} a^{- 3\gamma_{\rm B} (q+1)}
\label{pot_pcbc_g}
\end{equation}
The dependence of the scalar field on the scale factor is also
given up to a single quadrature: 
\begin{equation}
\phi(a) = \pm \sqrt{\frac{3\gamma_{\rm B}}
{4 \pi \ell_{\rm Pl}^2}}\int \frac{da}{a}
\sqrt{\frac{D (q+1)}{2-\frac{1}{3} \frac{d \ln D}{d \ln a}}}
\label{phi_pcbc_g}
\end{equation}
where in order to satisfy the positivity condition of $D$,
the following conditions
\beqn
6 &>& \frac{d \ln D}{d \ln a}\;\;\;\;\;{\rm for} \;\;q > -1 \nonumber\\
6 &<& \frac{d \ln D}{d \ln a}\;\;\;\;\;{\rm for} \;\;q < -1
\label{class_pcbc_g}
\eeqn
must be imposed. 

It is worth noting that Eqs.~(\ref{pot_pcbc_g}) and (\ref{phi_pcbc_g})
follow without imposing any relation between the evolution of
the field's kinetic and potential energies. 
Therefore, we can obtain an explicit form for 
the potential and the scalar field once the form of 
the correction function $D$ has been specified. 

\subsection{Case A: $DV/\dot{\phi}^2 = \beta = {\rm constant}$}
\label{subsec5a}

For $\beta > -1/2$, the kinetic correction function is given by 
Eq. (\ref{caseAcorr}): 
\begin{equation}
D(a)=D_c A^{1+2\beta} a^{6-3\gamma_{\rm B} (1+2\beta)(q+1)}
\label{d_pcbc_a}
\end{equation}
We obtain the potential as a function of the scale factor
after substituting Eq.~(\ref{d_pcbc_a}) into Eq.~(\ref{pot_pcbc_g}):
\begin{equation}
V(a) = \frac{2\beta}{1+2\beta} a^{-3\gamma_{\rm B} (q+1)}
\label{pot_pcbc_a1}
\end{equation}
which is, of course, consistent with Eq.~(\ref{potA}).
The evolution of the scalar field is also given by substituting
Eq.~(\ref{d_pcbc_a}) into Eq.~(\ref{phi_pcbc_g}):
\beqn
\phi (a) &=& \sqrt{D_{ci}} 
a^{3-\frac{3}{2}\gamma_{\rm B}(1+2\beta)(q+1)}\nonumber\\
&&\;\;\;\;\;{\rm for} \;\;q \neq 
\frac{2-\gamma_{\rm B}(1+2\beta)}{\gamma_{\rm B}(1+2\beta)}
\nonumber\\
\phi (a) &=& \sqrt{D_{ce}} \ln a\nonumber\\
&&\;\;\;\;\;{\rm for} \;\;
q =
\frac{2-\gamma_{\rm B}(1+2\beta)}{\gamma_{\rm B}(1+2\beta)}
\label{phi_pcbc_a}
\eeqn
where 
\beqn
D_{ci} &=& \frac{D_c A^{1+2\beta}}
{3 \pi (1+2\beta)\{2-\gamma_{\rm B}(1+2\beta)(q+1)\}^2 \ell_{\rm Pl}^2}
\nonumber\\
D_{ce} &=& \frac{3 D_c A^{1+2\beta}}{4\pi(1+2\beta) \ell_{\rm Pl}^2}
\eeqn

Finally, the dependence of the potential on the scalar field 
follows from Eqs.~(\ref{pot_pcbc_a1}) and (\ref{phi_pcbc_a}): 
\beqn
V(\phi) &=& \frac{2\beta}{1+2\beta} 
\left(\frac{\phi}{\sqrt{D_{ci}}}\right)
^{\frac{-2\gamma_{\rm B} (q+1)}
{2-\gamma_{\rm B} (1+2\beta)(q+1)}}\nonumber\\
&&\;\;\;\;\;{\rm for} \;\;q \neq 
\frac{2-\gamma_{\rm B}(1+2\beta)}{\gamma_{\rm B}(1+2\beta)}
\nonumber\\
V(\phi) &=& \frac{2\beta}{1+2\beta}  
\exp \left[-3\gamma_{\rm B} (q+1) \frac{\phi}{\sqrt{D_{ce}}}\right]
\nonumber\\
&&\;\;\;\;\;{\rm for} \;\;q =
\frac{2-\gamma_{\rm B}(1+2\beta)}{\gamma_{\rm B}(1+2\beta)}
\label{pot_pcbc_a2}
\eeqn

\subsection{Case B: $V/\dot{\phi}^2 = \epsilon = {\rm constant}$}

In this case, the form of the kinetic function can be obtained by 
substituting Eq.~(\ref{rho_to_n_pcbc_g}) into 
Eq.~(\ref{bernsol}): 
\begin{equation}
\frac{1}{D(a)} = 
\frac{2\epsilon \gamma_{\rm B} (q+1)}{2-\gamma_{\rm B} (q+1)}
+\frac{C_1}{A^{-1}}a^{-6+3\gamma_{\rm B} (q+1)}
\end{equation}
for $q \neq (2-\gamma_{\rm B})/\gamma_{\rm B}$ and 
\begin{equation}
\frac{1}{D(a)} = 12\epsilon \ln a + C_2\nonumber\\
\end{equation}
for $q = (2-\gamma_{\rm B})/\gamma_{\rm B}$, 
where $C_1$ and $C_2$ are integration constants.

\subsection{Case C: $D = D_c a^p$}

In the previous two cases, the corresponding loop--inspired cosmology
was obtained by first specifying a relation between the 
field's kinetic and potential energies, which as we have argued above is analogous 
to the condition for a scaling solution in conventional cosmology. 
However, for the scenario considered in this Section, the condition 
that $\gamma_{\rm L,eff}$ is constant simplifies the general 
analysis and leads to another way of obtaining the 
correspondent loop-inspired cosmology. This involves
specifying the form of the kinetic correction function 
$D(a )$, as indicated by the form of 
Eqs.~(\ref{pot_pcbc_g}) and (\ref{phi_pcbc_g}). This provides a
complementary approach which we now develop further with a 
particular example that allows for a direct comparison with 
the discussion of subsection \ref{subsec5a}. 

Let us consider the case 
\beq
D = D_c a^p
\label{d_pcbc_c}
\eeq
where $D_c$ and $p$ are positive constants.
This power-law behaviour arises as the asymptotic
limit of the LQC eigenvalue correction function (\ref{approxD}) 
in the semi-classical limit of LQC. 
Substitution of Eq.~(\ref{d_pcbc_c}) into Eq.~(\ref{pot_pcbc_g})
yields the potential in terms of the scale factor:
\beq
V(a) = \frac{6-p-3\gamma_{\rm B}}{6-p} a^{-3\gamma_{\rm B} (q+1)}
\label{pot_pcbc_c1}
\eeq
for $p \ne 6$ and substitution of Eq.~(\ref{d_pcbc_c}) 
into Eq.~(\ref{phi_pcbc_g}) determines the evolution of the scalar field as
\beq
\phi (a) = \pm \sqrt{\frac{9 \gamma_{\rm B} D_c}{4\pi \ell^2_{pl}}}
\sqrt{\frac{q+1}{6-p}} \frac{2}{p} a^{p/2}
\label{phi_pcbc_c}
\eeq
where an integration constant has been set to zero without loss
of generality by performing a linear shift
in the value of the scalar field. Eq.~(\ref{class_pcbc_g})
implies that if the correction function is 
to remain positive--definite, we require  
$p<6$ $(p>6)$ for $q>-1$ $(q<-1$).
%\label{class_pcbc_c}

We then deduce the form of the potential from 
Eqs.~(\ref{pot_pcbc_c1}) and (\ref{phi_pcbc_c}):
\beq
V(\phi) = \frac{6-p-3\gamma_{\rm B}(q+1)}{6-p} B_i 
\phi^{-6\gamma_{\rm B}(q+1)/p}
\label{pot_pcbc_c2}
\eeq
where 
\beq
B_i= \left(\frac{9 \gamma_{\rm B} D_c}{\pi \ell_{\rm Pl}^2 p^2}
\frac{q+1}{6-p}\right)^{3\gamma_{\rm B} (q+1) /p}
\eeq
and the positive root in Eq.~(\ref{phi_pcbc_c}) has been 
chosen without loss of generality. 
Comparing Eqs. (\ref{powerchi}) and (\ref{pot_pcbc_c2}), 
it is interesting that for this case the inverse power-law potential plays
a central role in realising $\gamma_{\rm eff} = {\rm constant}$
in {\em both} the braneworld and 
loop-inspired cosmologies. 

Finally, we consider the case $p=6$  for completeness.
Eq.~(\ref{gammaeffL}) implies that $\gamma_{\rm B} (q+1)=0$ and 
both $\gamma_{\rm B}=0$ and $q=-1$ result in a constant Hubble parameter, 
i.e., exponential expansion. The scalar field dynamics is then determined 
from Eq. (\ref{loopscalareom}) to be either $\dot{\phi}=0$ or 
$\dot{\phi}^2 \propto D (a) $ with a constant potential in both cases.  

\subsection{Case D: $D = 1$}

In order to recover conventional cosmology at low energies,
$D (a)$ should approach unity as the size of the universe increases.
If we choose this special form for the correction function, 
the dependence of the potential and scalar field can be read off 
immediately from Eqs.~(\ref{pot_pcbc_g}) and (\ref{phi_pcbc_g}): 
\beq
V(a) = \frac{2-\gamma_{\rm B}(q+1)}{2} a^{-3\gamma_{\rm B} (q+1)}
\label{pot_pcbc_d1}
\eeq
\beq
\phi(a) = \sqrt{\frac{3 \gamma_{\rm B} (q+1)}{8 \pi \ell_{\rm Pl}^2}}
\ln a
\label{phi_pcbc_d}
\eeq
where the allowed value of $q$ is constrained
as $q>-1$ since $2-(1/3)(d \ln D/ d \ln a)$ is always
positive for this case. The potential then follows
from Eqs.~(\ref{pot_pcbc_d1}) and (\ref{phi_pcbc_d}):
\beq
V(a) = \frac{2-\gamma_B(q+1)}{2}
\exp [-\sqrt{24\pi \ell_{\rm Pl}^2 \gamma_{\rm B}(q+1)}\phi]
\label{pot_pcbc_d2}
\eeq
In effect, by choosing $D=1$, we are expressing the braneworld 
dynamics in terms of an effective conventional model 
based on general relativity with an 
equation of state $\gamma_{\rm eff} = \gamma_{\rm B}(1+q)$. 

\section{Summary}

The aim of this paper has been to investigate 
the extent to which two apparently disparate approaches 
to early universe cosmology, namely the braneworld scenario and 
loop quantum cosmology, may share some common features, 
at least from a dynamical point of view. From a 
phenomenological perspective, both paradigms are characterized 
by the way in which they lead to modifications of the 
standard Friedmann equations of classical cosmology. 
It has been shown that within the context of a general (spatially curved) 
FRW background sourced by a scalar field, 
a dynamical correspondence can be established 
between a given braneworld model and 
a class of loop--inspired backgrounds, in the 
sense that both approaches lead to an identical Hubble expansion if 
the correction terms arising in the 
Friedmann equations are related in an appropriate way. 

This provides an alternative framework for 
regarding braneworld cosmology as a constrained Hamiltonian system. 
Alternatively, it was shown in \cite{Copeland:2004qe} how the general 
class of braneworld cosmologies
(\ref{branefriedmann}) can be reformulated in such a way that 
they take an identical form to that of the plane--autonomous 
system for a minimally coupled scalar field in 
standard, relativistic cosmology.  In principle, the correspondence we have 
established above could be employed to reformulate loop--inspired 
scenarios in a similar way. It would 
be interesting to consider this possibility further since it 
would enable a stability analysis to be readily performed. 
A further extension of our analysis 
would be to include a perfect fluid as well as a scalar 
field in the matter sector and investigate whether the above 
correspondences can be extended to this more general case \cite{param}.

Finally, we considered cases where the 
cosmological background represents a scaling solution. 
In particular, when corrections to the Friedmann equation 
take a power--law form in both the braneworld and loop--inspired pictures
(as is the case for example in a number of high--energy limits),  
the scaling solution in both paradigms 
is driven by a scalar field with a simple power--law potential. 
Hence, this provides a concrete example where the two
approaches lead to qualitatively similar behaviour.

%========================================%
%<<<<<<<<<< ACKNOWLEDGMENT >>>>>>>>>>>%
%========================================%
\section*{Acknowledgments}
SM is grateful to the University of Nottingham
and Queen Mary, University of London 
for their hospitality during a period when this work 
was initiated. He is supported by a Grant-in-Aid for
Scientific Research Fund of the Ministry of Education,
Culture, Sports, Science and Technology 
(Young Scientists (B) 17740154). EJC is grateful to the Aspen 
Center for Physics, for their hospitality during a period when some of 
this  work was being carried out. We thank P. Singh for helpful 
discussions and comments. 

%=================================

\end{document}